\documentclass[preprint,showpacs,preprintnumbers,amsmath,amssymb,nofootinbib]{revtex4}

\usepackage{graphicx}
\usepackage{dcolumn}
\usepackage{bm}
\usepackage{amsmath,dsfont}
\usepackage{color}

\def\IR{{\mathds{R}}}

\def\E{{\mathcal{E}}}

\def\H{{\mathcal{H}}}

\def\Q{{\mathcal{Q}}}
\def\U{{\mathcal{U}}}
\def\N{{\mathcal{N}}}
\def\C{{\mathcal{C}}}

\def\p{{\partial}}

\def\vp{{\vec{p}}}
\def\hx{{\hat{r}}}
\def\vn{{\hat{n}}}

\def\vJ{{\vec{J}}}
\def\vL{{\vec{L}}}
\def\vo{{\vec{\Omega}}}
\def\vK{{\vec{K}}}

\def\vA{{\vec{A}}}
\def\vB{{\vec{B}}}

\def\vG{{\vec{G}}}
\def\vC{{\vec{C}}}

\def\vpsi{{\vec{\psi}}}
\def\vxi{{\vec{\xi}}}

\def\vsigma{{\vec{\sigma}}}

\def\smallover#1/#2{\hbox{$\textstyle\frac{#1}{#2}$}}
\def\IR{{\mathds{R}}}

\def\M{{\mathcal{M}}}

\def\b{{\beta}}
\def\Id{{1\!\!1}}
\def\a{{\alpha}}
\def\vA{{\vec{A}}}
\def\vK{{\vec{K}}}
\def\vJ{{\vec{J}}}
\def\vPi{{\vec{\Pi}}}

\def\gyro{{g}}
\def\b{{\beta}}
\def\d{{\delta}}

\def\vp{{\vec{p}}}
\def\vx{{\vec{x}}}

\def\vn{{\vec{n}}}

\def\vB{{\vec{B}}}
\def\vS{{\vec{S}}}
\def\vnabla{{\vec{\nabla}}}
\def\hx{{\frac{\vec{x}}{r}}}

\def\beq{\begin{equation}}
\def\eeq{\end{equation}}
\def\beqa{\begin{eqnarray}}
\def\eeqa{\end{eqnarray}}

\def\osp{{\mathfrak{osp}}}

\begin{document}

\preprint{arxiv:1003.0137}

\title{Dynamical supersymmetry of spin particle-magnetic field interaction}

\author{\large J.-P.~Ngome\textsuperscript{a}}
\author{\large P.~A.~Horv\'athy\textsuperscript{a}}
\author{\large J.~W.~van Holten\textsuperscript{b}}
\affiliation{\textsuperscript{a}
Laboratoire de Math\'ematiques et de Physique Th\'eorique, 
Universit\'e Fran\c cois-Rabelais Tours,
F\'ed\'eration Denis Poisson - CNRS
Parc de Grandmont, 37200 Tours, France.
}
\email{juste.ngome-at-lmpt.univ-tours, horvathy-at-lmpt.univ-tours.fr}
\affiliation{\textsuperscript{b}
NIKHEF, PO Box 41882, 1009 DB Amsterdam, Netherlands.
}

\email{t32-at-nikhef.nl}

\begin{abstract}
We study the super and dynamical symmetries of a fermion  in a monopole background. The Hamiltonian also involves an additional spin-orbit coupling term, which is parameterized by the gyromagnetic ratio. We construct the superinvariants associated with the system using a SUSY extension of a previously proposed  algorithm, based on Grassmann-valued Killing tensors. Conserved quantities arise for certain definite values of the gyromagnetic factor~:
$\N=1$ SUSY requires $g=2$; a Kepler-type dynamical symmetry only arises, however,
for the anomalous values $g=0$  and $g=4$. The two anomalous systems can be unified into an $\N=2$ SUSY system built by doubling the number of Grassmann variables. The planar system also exhibits an $\N=2$ supersymmetry without Grassmann variable doubling.
\end{abstract}


\maketitle

\section{Introduction}
\noindent
Following a classical result of
D'Hoker and Vinet  \cite{DV1} (see also  \cite{ GvH,DJ-al,HMH,HmonRev,Pl}) a non-relativistic spin-$\frac{1}{2}$ charged particle with  gyromagnetic ratio $\gyro=2$, interacting with a point magnetic monopole, admits an $\osp(1|2)$ 
supersymmetry.
It has no Runge-Lenz type
dynamical symmetry, though \cite{FeherO31}.

Another, surprising, result of D'Hoker and Vinet \cite{DV2} says, however, that a 
non-relativistic spin-$\frac{1}{2}$ charged particle with \emph{anomalous gyromagnetic ratio} $\gyro=\,4\,$, interacting with a point magnetic monopole plus a Coulomb plus a fine-tuned inverse-square potential, 
 does have such a dynamical symmetry.
  This  is to be compared with the one 
 about the ${\rm O}(4)$ symmetry of a scalar particle
 in such a combined field \cite{MICZ}.
Replacing the scalar particle by a spin $1/2$ particle
with gyromagnetic ratio $\gyro=0$, one can prove that
two anomalous systems, the one with $\gyro=4$ and the one with $\gyro=0$ are, in fact, superpartners \cite{FH}. Note that
for both particular $\gyro$-values, one also 
has an additional ${\rm o}(3)$ ``spin'' symmetry.

On the other hand, it has been shown by Spector \cite{Spector} that the $\,\N=1\,$ supersymmetry only allows $g=2$ and no scalar potential. Runge-Lenz and SUSY appear, hence, inconsistent.

In this paper, we investigate the bosonic as well as supersymmetries of the Pauli-type Hamiltonian,
\beq
\H_{\gyro}=\frac{\vPi^2}{2}-\frac{e\gyro}{2}\,\vS\cdot\vB+V(r)\,,\label{Hamiltonian}
\eeq
which describes the motion of a fermion with spin $\,\vS\,$ and electric charge $\,e\,$, in  the combined magnetic field, $\,\vB\,$, plus a spherically symmetric scalar field $V(r)$, which also includes a Coulomb term (a ``dyon'' in what follows).
In (\ref{Hamiltonian}), $\,\vPi=\vp-e\,\vA\,$ denotes the gauge covariant momentum and the constant $\,\gyro\,$ represents the gyromagnetic ratio of the spinning particle.
Except in Section \ref{S7}, the gauge field is taken that of an Abelian monopole.

We derive the (super)invariants by considering the Grassmannian extension of the algorithm proposed before by one of us \cite{vH}.

The main ingredients are Killing tensors, determined by a linear system of first order partial differential equations. 
 
Our recipe has already been used successfully to derive bosonic symmetries \cite{vH,H-N,Ngome,Visi}; in this paper we systematically extend these results to supersymmetries associated with Grassmann-algebra valued Killing tensors  
\cite{vH,vH-al,GvH,Visi2}.

The plan of this paper is as follows~: in Section \ref{S2} we derive the equations of the motion of the system. In Section \ref{S3}, we present the general formalism and we analyse the conditions under which conserved quantities are generated. In Sections \ref{S4} and \ref{S5} we investigate the super resp.  bosonic symmetries of 
the fermion-monopole system. Our investigations confirm Spector's theorem.

In section \ref{S6}, we show, however, that the obstruction can be overcome by 
a \emph{dimensional extension of  fermionic space} \cite{Salomonson,Michelson,A-K}. Working with two, rather than just one
Grassmann variable allows us to combine the two anomalous systems into one with $\,\N=2\,$ supersymmetry. In section \ref{S7}, we investigate the SUSY of the spinning particle coupled with a static magnetic field in the plane.

\section{Hamiltonian Dynamics of the spinning system}\label{S2}


\noindent
Let us consider a charged spin-$\frac{1}{2}$ particle moving in a flat manifold $\,\M^{D+d}\,$ which is the extension of the bosonic configuration space $\,\M^{D}\,$ by a $d$-dimensional internal space carrying the fermionic degrees of freedom. The $(D+d)$-dimensional space $\,\M^{D+d}\,$ is described by the local coordinates $\left(\,x^{\mu},\,\psi^{a}\right)$ where $\,\mu=1,\cdots,D\,$ and $\,a=1,\cdots,d\,$. The motion of the spin-particle  is, therefore, described by the curve $\tau\rightarrow\left(\,x(\tau),\,\psi(\tau)\right)\,\in\,\M^{D+d}\,$. We choose $\,D=d=3\,$ and we focus our attention to the spin-$\frac{1}{2}$ charged particle interacting with the static $\,U(1)\,$ monopole background, $\,\displaystyle{\vB=\vnabla\times\vA=\frac{q}{e}(\vx/r^3)}\,$, such that the system is described by the Hamiltonian (\ref{Hamiltonian}). In order to deduce, in a classical framework, the supersymmetries and conservation laws, we introduce the covariant hamiltonian formalism, with basic phase-space variables $\,\left(x^{j},\Pi_{j},\psi^{a}\right)\,$. Here the variables $\,\psi^{a}\,$ transform as tangent vectors and satisfy the Grassmann algebra, $\,\psi^{i}\psi^{j} + \psi^{j}\psi^{i}=0\,$. The internal angular momentum of the particle 
can also be described
in terms of vector-like Grassmann variables,
\beq
S^{j}=-\frac{i}{2}\epsilon^j_{\,\,kl}\psi^{k}\,\psi^{l}\,.
\eeq
Defining the covariant Poisson-Dirac brackets for functions $\,f\,$ and $\,h\,$ of the phase-space as
\beqa
\big\{f,h\big\}&=&\p_j f\,\frac{\p h}{\p \Pi_j}-\frac{\p f}{\p \Pi_j}\,\p_j h 
+eF_{ij}\,\frac{\p f}{\p \Pi_i}\frac{\p h}{\p \Pi_j}
+i(-1)^{a^{f}}\frac{\p f}{\p \psi^a}\frac{\p h}{\p \psi_{a}}\,,\label{PBrackets}
\eeqa
where $\,a^{f}=\left(0,1\right)\,$ is the Grassmann parity of the function $\,f$ and the magnetic field reads $\,B_{i}=(1/2)\epsilon_{ijk}F_{jk}\,$. It is straightforward to obtain the non-vanishing fundamental brackets,
\beqa
\big\{x^{i},\,\Pi_{j}\big\}=\d^{i}_{j},\quad\big\{\Pi_{i},\,\Pi_{j}\big\}=e\,F_{ij},\quad\big\{\psi^{i},\,\psi^{j}\big\}=-i\,\d^{ij}\,,\\[8pt]
\big\{S^{i},\,G^{j}\big\}=\epsilon^{\;\,ij}_{k}\,G^{k}\quad\hbox{with}\quad G^{k}=\psi^{k},\,S^{k}\,.
\eeqa
It follows that, away from the monopole's  location,  the Jacobi identities are verified \cite{Jackiw84,Chaichian}.

The equations of the motion  can  be obtained in this covariant Hamiltonian framework \footnote{The dot means derivative w.r.t. the evolution parameter $\,\frac{d}{d\tau}\,$.},
\beqa
\dot{\vG}=
\frac{e\gyro}{2}\,\vG\times\vB\,,
\label{EqM}\\[6pt]
\dot{\vPi}=
e\,\vPi\times\vB-\vnabla{V(r)}+\frac{e\gyro}{2}\,\vnabla{\left(\vS\cdot\vB\right)}\,.\label{Lorentz}
\eeqa
Equation (\ref{EqM}) shows  that the fermionic vectors $\,\vS\,$ and $\,\vpsi\,$ are conserved when the spin and the magnetic field are uncoupled, i.e. for \emph{vanishing gyromagnetic ratio}, $\,\gyro=0\,$.  Note that, in addition to the magnetic field term, the Lorentz equation (\ref{Lorentz}) also involves a potential term augmented  with a spin-field interaction term.

\section{Killing tensors for fermion-monopole system}\label{S3}
\noindent
Now we outline the algorithm developed in \cite{vH} to
construct constants of the motion. First, a phase-space function associated with a (super)symmetry can be expanded in powers of the covariant momenta,
\beq
\Q\left(\vx,\,\vPi,\,\vpsi\right)=C(\vx,\,\vpsi)+\sum_{k=1}^{p-1}\,\frac{1}{k!}\,C^{i_1\cdots i_k}(\vx,\,\vpsi)\,\Pi_{i_1}\cdots\Pi_{i_k}\,.
\label{Exp}
\eeq
Requiring that $\Q$ Poisson-commutes with the Hamiltonian,
$\big\{\H_{\gyro},\Q\big\}=0\,,$
implies the series of constraints,  
\beq
\begin{array}{llll}\displaystyle{
C^{i}\,\p_i V+\frac{ie\gyro}{4}\,\psi^l\psi^m C^j\,\p_j F_{lm}-\frac{e\gyro}{2}\,\psi^m\frac{\p C}{\p\psi^a}\,F_{am}
=0},&\hbox{order 0}
\\[10pt]\displaystyle{
\p_jC=C^{jk}\p_{k}V+eF_{jk}C^k+\frac{ie\gyro}{4}\psi^l\psi^m C^{jk}\p_k F_{lm}-\frac{e\gyro}{2}\psi^m\frac{\p C^j}{\p\psi^a}F_{am}},
&\hbox{order 1}
\\[11pt]\displaystyle{
\p_jC^k+\p_kC^j=C^{jkm}\p_{m}V+e\left(F_{jm}C^{mk}+F_{km}C^{mj}\right)+\frac{ie\gyro}{4}\psi^l\psi^m C^{ijk}\p_i F_{lm}-\frac{e\gyro}{2}\psi^m\frac{\p C^{jk}}{\p\psi^a}F_{am}},
&\hbox{order 2}\\[16pt]
\p_jC^{kl}+\p_lC^{jk}+\p_kC^{lj}= C^{jklm}\p_{m}V+e\left(F_{jm}C^{mkl}+F_{lm}C^{mjk}+F_{km}C^{mlj}\right)
\\[8pt]
\quad\qquad\qquad\qquad\qquad\qquad\qquad\qquad\quad+\;\displaystyle{\frac{ie\gyro}{4}\psi^m\psi^n C^{ijkl}\p_i F_{mn}-\frac{e\gyro}{2}\psi^m\frac{\p C^{jkl}}{\p\psi^a}F_{am}}\,,
&\hbox{order 3}
\\
\vdots\qquad\qquad\qquad\qquad\qquad\vdots&\vdots
\end{array}
\label{constraints}
\eeq
This series can be truncated at a finite order, \textit{p}, provided  \textit{the constraint of order
$p$ becomes a Killing equation}. The zeroth-order equation can be interpreted as a  \textit{consistency condition between the potential and the (super)invariant}. Apart from the zeroth-order constants of the motion, i.e., such that do not depend on the  momentum, all other order-\textit{n} (super)invariants are deduced by the systematic method (\ref{constraints}) implying rank-\textit{n} Killing tensors. Each Killing tensor solves the higher order constraint of (\ref{constraints}) and can generate a conserved quantity.

 In this paper, we are interested by (super)invariants which are linear or quadratic in the momenta.
Thus, we have to determine generic Grassmann-valued Killing tensors of rank-one and rank-two.
 
$\bullet$ Let us first investigate the Killing equation,
\beq
\p_jC^{k}(\vx,\,\vpsi)+\p_kC^{j}(\vx,\,\vpsi)=0\,.\label{KConstraint1}
\eeq
Following Berezin and Marinov \cite{B-M}, any tensor which takes its values in the Grassmann algebra may be represented as a finite sum of homogeneous monomials,
\beq
C^{i}(\vx,\,\vpsi)=\sum_{k\geq 0}\C^{i}_{a_{1}\cdots a_{k}}(\vx)\psi^{a_{1}}\cdots\psi^{a_{k}}\,,\label{ExternalTensor}
\eeq
where the coefficients tensors, $\C^{i}_{a_{1}\cdots a_{k}}$, are completely anti-symmetric in the fermionic indices $\,\left\lbrace a_{k}\right\rbrace\,$. The tensors (\ref{ExternalTensor}) satisfy (\ref{KConstraint1}), from which we deduce that their (tensor) coefficients  satisfy,
\beq
\p_j\C^{k}_{a_{1}\cdots a_{k}}(\vx)+\p_k\C^{j}_{a_{1}\cdots a_{k}}(\vx)=0\quad\Longrightarrow\quad\p_{i}\p_{j}\C^{k}_{a_{1}\cdots a_{k}}(\vx)=0\,,\label{KConstraint1bis}
\eeq
providing us with the most general rank-1 Grassmann-valued Killing tensor
\beq
C^{i}(\vx,\,\vpsi)=\sum_{k\geq 0}\big(M^{ij}\,x^{j}+N^{i}\big)_{a_{1}\cdots a_{k}}\psi^{a_{1}}\cdots\psi^{a_{k}}\,,\quad M^{ij}=-M^{ji}\,.\label{Exp2}
\eeq
Here $\,N^{i}\,$ and the antisymmetric $\,M^{ij}\,$ are constant tensors.

$\bullet$ Let us now construct the rank-2 Killing tensors which solve the Killing equation,
\beq
\p_jC^{kl}(\vx,\,\vpsi)+\p_lC^{jk}(\vx,\,\vpsi)+\p_kC^{lj}(\vx,\,\vpsi)=0\,.\label{KConstraint2}
\eeq
We consider the expansion in terms of Grassmann degrees of freedom \cite{B-M} and the coefficients $\,\C ^{ij}_{a_{1}\cdots a_{k}}\,$ are constructed as symmetrized products \cite{G-R} of  Yano-type Killing tensors, $\C^{i}_{\,Y}(\vx)\,$, associated with the rank-1 Killing tensors $\,\C ^{i}(\vx)\,$,
\beq
\C^{ij}_{a_{1}\cdots a_{k}}(\vx)=\frac{1}{2}\left(\C^{i}_{\,Y}\widetilde{\C}^{jY}+\widetilde{\C}^{i}_{\,Y}\C^{jY}\right)_{a_{1}\cdots a_{k}}\,.\label{Coef}
\eeq
It is worth noting that the Killing tensor (\ref{Coef}) is symmetric in its bosonic indices and anti-symmetric in the fermionic indices. Thus, we obtain
\beq
C^{ij}(\vx,\,\vpsi)=\sum_{k\geq 0}\left(
M^{(i}_{\;ln}\widetilde{M}^{j)\,n}_{\;m}x^lx^m+M^{(i}_{\;ln}\widetilde{N}^{j)\,n}x^l+N^{(i}_{\;n}\widetilde{M}^{j)\,n}_{\;m}x^m+N^{(i}_{\;n}\widetilde{N}^{j)\,n}\right)_{a_{1}\cdots a_{k}}\psi^{a_{1}}\cdots\psi^{a_{k}}\,,\label{Exp3}
\eeq
where $\,M^{ij}_{\;\;k}\,$, $\,\widetilde{M}^{ij}_{\;\;k}\,$, $\,N^{j}_{\;k}\,$ and $\,\widetilde{N}^{j}_{k}\,$ are skew-symmetric constants tensors. Then one can verify with direct calculations that (\ref{Exp2}) and (\ref{Exp3}) satisfy the Killing equations.

\section{SUSY of fermion in magnetic monopole field}\label{S4}

\noindent
Having constructed the generic Killing tensors (\ref{Exp2}) and (\ref{Exp3}) generating constants of the motion, we now describe the supersymmetries of the Pauli-like Hamiltonian (\ref{Hamiltonian}). To start, we search for momentum-independent invariants, i.e. which are not derived from a Killing tensor, $\,C^{i}= C^{ij}= \cdots = 0\,$. In this case, the system of equations (\ref{constraints}) reduces to the constraints,
\beqa\left\lbrace
\begin{array}{lll} \displaystyle{
\gyro\psi^m\frac{\p \Q_{0}(\vx,\vpsi)}{\p\psi^a}\,F_{am} =0}\,,\qquad &\hbox{order 0}&
\\[12pt]\displaystyle{
\p_i\Q_{0}(\vx,\vpsi)=0}&\hbox{order 1}\,.&
\end{array}
\right.
\eeqa
For $\gyro=0$ [which means no spin-gauge field coupling], it is straightforward to see that the spin vector, together with an arbitrary function $f(\vpsi\,)$ which only depends on the Grassmann variables, are conserved.

\textit{For nonvanishing gyromagnetic ratio $\gyro$, only the ``chiral" charge} $\Q_{0}=\vpsi\cdot\vS\,$ remains  conserved. Hence, the charge $\,\Q_{0}\,$ can be considered as the projection of the internal angular momentum, $ \vS$, onto the internal trajectory $\,\psi(\tau)\,$. Thus $\Q_{0}\,$ can be viewed as the internal analogue of the projection of the angular momentum, in bosonic sector, onto the classical trajectory $\,x(\tau)\,$.

Let us now search for superinvariants which are linear in the covariant momentum. $C^{ij}=C^{ijk}=\cdots =0\,$ such that (\ref{constraints}) becomes
\beqa\left\lbrace \begin{array}{lll} 
 \displaystyle{ C^{i}\,\p_i V+\frac{ie\gyro}{4}\,\psi^l\psi^m C^j(\vx,\vpsi)\,\p_j F_{lm}-\frac{e\gyro}{2}\psi^m\frac{\p C(\vx,\vpsi)}{\p\psi^a}\,F_{am}
=0}\,,&\hbox{order 0}& 
\\[7pt]
\displaystyle{
\p_jC(\vx,\vpsi)=eF_{jk}C^k(\vx,\vpsi)-\frac{e\gyro}{2}\psi^m\frac{\p C^j(\vx,\vpsi)}{\p\psi^a}\,F_{am}}\,, &\hbox{order 1}& 
\\[12pt]
\p_jC^k(\vx,\vpsi)+\p_kC^j(\vx,\vpsi)=0&\hbox{order 2}\,.&
\label{AngMom}
\end{array}\right.
\eeqa
We choose the non-vanishing $\,N^{j}_{a}=\d^{j}_{a}\,$ in the general rank-1 Killing tensor (\ref{Exp2}).
This provides us with the rank-1 Killing tensor generating the supersymmetry transformation, 
\beq
\,C^j(\vx,\vpsi)=\d^{j}_{a}\,\psi^{a}\,.
\label{SUSYKILLING}
\eeq
By substitution of this Grassmann-valued Killing tensor into the first-order equation of (\ref{AngMom}) we get 
\beq
\vnabla C(\vx,\vpsi)=\frac{q}{2}\left(\gyro-2\right)\frac{\vx\times\vpsi}{r^3}\,.\label{PrExp}
\eeq
Consequently, a solution 
$C(\vx,\vpsi)=0$ of (\ref{PrExp}) is only obtained for a fermion with ordinary gyromagnetic ratio 
\beq
\gyro=2\,.
\label{g2}
\eeq
Thus we obtain, for $\,V(r)=0\,$, the \textit{Grassmann-odd supercharge} generating the $\N=1$ supersymmetry of the spin-monopole field system,
\beq
\Q=\vpsi\cdot\vPi\,,
\qquad
\big\{\Q,\,\Q\big\}=-2i\H_{2}\,.
\label{SC0}
\eeq
 
For nonvanishing potential, $V(r)\neq 0\,$, the zeroth-order consistency condition of (\ref{AngMom}) is expressed as \footnote{We use the identity 
$\,S^kG^j\p_j B^k=\psi^l\psi^m\,G^j\p_j F_{lm}=0\,$.}
$
({V'(r)}/{r})\,\vpsi\cdot\vx=0\,.$ Consequently, adding \emph{any} spherically symmetric potential $V(r)\,$ breaks the supersymmetry  generated by the Killing tensor $C^j=\d^{j}_{a}\,\psi^{a}$~:
$\N=1$ SUSY requires an ordinary gyromagnetic factor, and no additional radial potential 
is allowed \cite{Spector}.

Another Killing tensor (\ref{Exp2}) is obtained by considering the particular case with the non-null tensor $ \,N^{j}_{\;a_1a_2}=\epsilon^{j}_{\;a_1a_2}\,
$. This leads to the rank-1 Killing tensor, 
\beq
C^j(\vx,\vpsi)=\nobreak\epsilon^{j}_{\;ab}\psi^{a}\psi^{b}\,.
\eeq
The first-order constraint of (\ref{AngMom}) is solved with $\,C(\vx,\vpsi)=0\,$, provided the gyromagnetic ratio takes the value $\,\gyro=2\,$. For vanishing potential, it is straightforward to verify the zeroth-order consistency constraint and to obtain \textit{the Grassmann-even supercharge},
\beq
\Q_{1}=\vS\cdot\vPi\,,\label{evenS1}
\eeq
\textit{defining the ``helicity" of the spinning particle}. As expected, the consistency condition of superinvariance under (\ref{evenS1}) is again violated for $\,V(r)\neq 0\,$, breaking the supersymmetry of the Hamiltonian $\,\H_{2}\,$, in (\ref{SC0}).

Let us now consider the rank-1 Killing vector, 
\beq
C^{j}(\vx,\vpsi)=\big(\vS\times\vx\big)^{j}\,,
\eeq 
 obtained by putting
$M^{ij}_{\;a_1a_2}=(i/2)\epsilon^{kij}\,\epsilon_{ka_1a_2}$ into the generic rank-1 Killing tensor (\ref{Exp2}). The first-order constraint 
is satisfied with $\,C(\vx,\vpsi)=0\,$, provided the particle carries  gyromagnetic ratio $\gyro=2$. Thus, we obtain the supercharge,
\beq 
\Q_{2}=(\vx\times\vPi)\cdot\vS\,,\label{q2}
\eeq
which, just like those in (\ref{SC0}) and (\ref{evenS1}) only appears when the potential is absent, $V=0$.

We consider the SUSY given when 
$\, M^{ij}_{\;a}=\epsilon^{\;\,ij}_{a}\,
$ so that the Killing tensor (\ref{Exp2}) reduces to 
\beq
C^{j}(\vx,\vpsi)=-\epsilon^{j}_{\;ka}x^{k}\psi^{a}\,.
\eeq 
The first-order constraint of (\ref{AngMom}) is solved with  $\,\displaystyle{C(\vx,\vpsi)=\frac{q}{2}\left(\gyro-2\right)\frac{\vpsi\cdot\vx}{r}}\,$. The zeroth-order consistency condition is, in this case, identically satisfied for an arbitrary  radial potential. We have thus constructed the Grassmann-odd supercharge,
\beq
\Q_{3}=(\vx\times\vPi)\cdot\vpsi+\frac{q}{2}\left(\gyro-2\right)\frac{\vpsi\cdot\vx}{r}\,,
\eeq
which is still conserved for a particle carrying an arbitrary gyromagnetic ratio $\gyro\,$; see also \cite{DJ-al}.

Now we  turn to superinvariants which are quadratic in the covariant momentum. For this, we solve the reduced series of constraints,
\beqa\left\lbrace
\begin{array}{llll}\displaystyle{C^{i}\p_i V+\frac{ie\gyro}{4}\,\psi^l\psi^m C^j\p_j F_{lm}-\frac{e\gyro}{2}\psi^m\frac{\p C}{\p\psi^a}F_{am}
=0},&\hbox{order 0}
\\[8pt]\displaystyle{
\p_jC=C^{jk}\p_k V+eF_{jk}C^k+\frac{ie\gyro}{4}\psi^l\psi^m\,C^{jk}\p_k F_{lm}-\frac{e\gyro}{2}\psi^m\frac{\p C^{j}}{\p\psi^a}\,F_{am}},
&\hbox{order 1}
\\[8pt]\displaystyle{
\p_jC^k+\p_kC^j=e\left(F_{jm}C^{mk}+F_{km}C^{mj}\right)-\frac{e\gyro}{2}\psi^m\frac{\p C^{jk}}{\p\psi^a}\,F_{am}}\,,
&\hbox{order 2}\\[8pt]\displaystyle{
\p_{j}C^{km}+\p_{m}C^{jk}+\p_{k}C^{mj}=0} &\hbox{order 3}\,.
\end{array}\right.
\label{RLV}
\eeqa
We first observe that $\,C^{ij}(\vx,\vpsi)=\d^{ij}$ is a constant Killing tensor. Solving the second- and the first-order constraints of (\ref{RLV}), we obtain $\,C^{j}(\vx,\vpsi)=0\,$ and $\,\displaystyle{C(\vx,\vpsi)= V(r)-\frac{e\gyro}{2}\vS\cdot\vB}\,$, respectively. The zeroth-order consistency condition is identically satisfied and we obtain the energy of the spinning particle,
\beq
\E= \frac{1}{2}\vPi^{2}-\frac{e\gyro}{2}\vS\cdot\vB+V(r)\,.
\eeq
 
Next, introducing the nonvanishing constants tensors,
$\,
M^{ijk}\!=\!\epsilon^{ijk}\,,\;\widetilde{N}^{ij}_{\;\,a}\!=\!-\epsilon_{\;\,a}^{ij}\,$, into (\ref{Exp3}), we derive the rank-2 Killing tensor with the property,
\beq
C^{jk}(\vx,\vpsi)=2\,\d^{jk}(\vx\cdot\vpsi)-x^j\psi^k-x^k\psi^j\,.\label{KTSUSY}
\eeq
Using the Killing tensor (\ref{KTSUSY}), we solve the second-order constraints of (\ref{RLV}) with 
$
\vC(\vx,\vpsi)=(q/2)\left(2-\gyro\right)\left(\vpsi\times\vx\right)/r\,.
$ 
 In order to deduce the integrability condition of the first-order constraint of (\ref{RLV}), we require the vanishing of the commutator,
\beq
\left[\p_{i},\,\p_{j}\right]C(\vx)=0\;\Longrightarrow\;\Delta\left( V(r)-\left(2-\gyro\right)^2\frac{q^2}{8r^2}\right)=0\,.\label{Laplace0}
\eeq
Then the Laplace equation (\ref{Laplace0}) provides us with the \textit{most general form of the potential admitting a Grassmann-odd supercharge quadratic in the velocity}, namely with
\beq
\displaystyle{V(r)=\left(2-\gyro\right)^2\frac{q^2}{8r^2}+\frac{\a}{r}+\b}\,.\label{PotSUSY}
\eeq 
Thus, we solve the first-order constraint with 
\beq
C(\vx,\vpsi)=\left(\frac{\a}{r}-e\gyro\vS\cdot\vB\right)\vx\cdot\vpsi\,,\label{NewSUSY}
\eeq
so that the zeroth-order consistency constraint is identically satisfied. Collecting our results leads to the Grassmann-odd supercharge quadratic in the velocity, 
\beq
\Q_{4}=\left(\vPi\times(\vx\times\vPi)\right)\cdot\vpsi+\frac{q}{2}\left(2-\gyro\right)\frac{\vx\times\vPi}{r}\cdot\vpsi +\left(\frac{\a}{r}-e\gyro\vS\cdot\vB\right)\vx\cdot\vpsi\,.
\eeq

This supercharge is \emph{not} a square root of the Hamiltonian $\,\H_g\,$, and that $\,\Q_4\,$ is conserved without restriction on the gyromagnetic factor, $\gyro\,$. We can also remark that for $\,\gyro=0\,$, the supercharge coincides with the scalar product of the \textit{separately conserved Runge-Lenz vector for a scalar particle \cite{MICZ} by the Grassmann-odd vector}:
\beq
\left. \Q_{4} \right|_{g = 0} = \vK_{s = 0}\cdot\vpsi\,.
\eeq 

The supercharges $\,\Q\,$ and $\,\Q_j\,$ with $\,j= 0,\cdots,3\,$, previously determined, form together, for ordinary gyromagnetic ratio, the classical superalgebra,
\beqa\begin{array}{lll}\displaystyle{
\big\{\Q_0,\,\Q_0\big\}=\big\{\Q_0,\,\Q_1\big\}=\big\{\Q,\,\Q_1\big\}=\big\{\Q_1,\,\Q_1\big\}=\big\{\Q_2,\,\Q_2\big\}=0}\,,\\[10pt]\displaystyle{

\big\{\Q_0,\,\Q\big\}=i\Q_{1}\,,\quad\big\{\Q_0,\,\Q_2\big\}=\big\{\Q_2,\,\Q_3\big\}=0}\,,\\[10pt]\displaystyle{

\big\{\Q_0,\,\Q_3\big\}=i\Q_{2}\,,\quad
\big\{\Q,\,\Q\big\}=-2i\H_{2}}\,,\\[10pt]\displaystyle{

\big\{\Q,\,\Q_2\big\}=\big\{\Q_1,\,\Q_3\big\}=\Q_4}\,,\\[10pt]\displaystyle{

\big\{\Q,\,\Q_3\big\}=2i\Q_1\,,\quad\big\{\Q_1,\,\Q_2\big\}=i\Q_3\Q\,,\quad\big\{\Q_3,\,\Q_3\big\}=i\left(2\Q_{2}-\Q_5\right)}\,,
\end{array}
\eeqa
where $\Q_5$ is the supercharge constructed in section \ref{S5}, cf. (\ref{Msquare}). From these results it follows, that the linear combination
$\Q_Y = \Q_3 - 2 \Q_0$ has the special property that its bracket with the standard supercharge $\Q$ vanishes:
\beq
\big\{ \Q_Y, \Q \big\} = 0.
\eeq
Indeed, $\Q_Y$ is precisely the Killing-Yano supercharge constructed in \cite{DJ-al}.

\section{Bosonic symmetries of the spinning particle }\label{S5}
\noindent
Let us investigate the bosonic symmetries of the Pauli-like Hamiltonian (\ref{Hamiltonian}). We use the generic Killing tensors constructed in section \ref{S3} to derive the associated constants of the motion. Firstly, we describe the rotationally invariance of the system by solving the reduced series of constraints (\ref{AngMom}).
For this, we consider the Killing vector provided by the replacement,
$\,
M^{ij}=-\epsilon^{ij}_{\;\;k}n^{k}\,
$ 
into  (\ref{Exp2}). Thus we obtain for any unit vector $\,\vn\,$, the generator of space rotations around $\,\vn\,$,
\beq
\vC(\vx,\vpsi)=\vn\times\vx\,.\label{KillingV} 
\eeq
Inserting the previous Killing vector in the first-order equation of (\ref{AngMom}) yields
$
C(\vx,\vpsi)=-q\,\left(\vn\cdot\vx\right)/r+c(\vpsi)\,.
$ 
The zeroth-order consistency condition of (\ref{AngMom}) requires, for arbitrary radial potential, $\,c(\vpsi)=\vS\cdot\vn\,$. Collecting our results provides us with the total angular momentum [which is plainly conserved for arbitrary gyromagnetic ratio],
\beq
\displaystyle{\vJ=\vL+\vS=
\vx\times\vPi-q\,\hx+\vS}\,.
\label{AngMomentum}
\eeq
In addition to the typical monopole term, 
${\vJ}$ also involves the spin vector, $\,\vS\,$. It
 generates an $o(3)_{rotations}$ bosonic symmetry algebra, $\,\big\{J^{i},J^{j}\big\}=\epsilon^{ijk}J^{k}$.

 In the case of vanishing gyromagnetic factor
 $\gyro=0$, the orbital part $\vL\,$ and the spin angular momentum $\vS\,$ are separately conserved involving an $o(3)_{rotations}\oplus o(3)_{spin}$ symmetry algebra.

Now we turn to invariants which are quadratic in the velocity. Then, we have to solve the series of constraints (\ref{RLV}). We first observe that for $\,M^{jmk}\!=\!\widetilde{M}^{jmk}\!=\!\epsilon^{jmk}\,$,
the Killing tensor (\ref{Exp3}) reduces to the rank-2 Killing-St\"ackel tensor,
\beq
C^{ij}(\vx,\vpsi)=2\d^{ij}\,\vx^{\,2}-2x^{i}x^j\,.\label{Momsquare}
\eeq
Inserting (\ref{Momsquare}) into the second- and in the first-order constraints of (\ref{RLV}), we get, for any gyromagnetic factor and for any arbitrary radial potential,
\beq
\vC(\vx,\vpsi)= 0\quad\hbox{and}\quad C(\vx,\vpsi)=-\gyro q\,\frac{\vx\cdot\vS}{r}\,.
\eeq
Thus, we obtain the Casimir,
\beq
\Q_{5}=\vJ^{2}-q^2+\left(\gyro-2\right)\vJ\cdot\vS-\gyro\Q_{2}\,,\label{Msquare}
\eeq
The bosonic supercharge $\,\Q_{5}$ is, as expected, \textit{the square of the total angular momentum, augmented with another, separately conserved term}. Indeed, for $\gyro=0\,$, it is straightforward to see that the spin
and hence
 $\,\vJ\cdot\vS \,$ are separately  conserved. For $\gyro=2\,$, we recover the conservation of $\Q_{2}$, cf.(\ref{q2}). For the anomalous gyromagnetic ratio  $\gyro=4\,$ we obtain that $\,\vJ\cdot\vS-2\Q_{2}\,$ is a constant of the motion.

Now we are interested by the hidden symmetry generated by  conserved Laplace-Runge-Lenz-type vectors, therefore we introduce into the algorithm (\ref{RLV}) the generator,
\begin{eqnarray}\displaystyle{ C^{ij}(\vx,\vpsi)= 2\,\d^{ij}\,\vn\cdot\vx-n^{i}x^j-n^{j}x^i}\,,
\label{RL1}
\end{eqnarray}
easily obtained by choosing the non-vanishing,
$\,\widetilde{N}^{ij}\!=\!\epsilon^{imj}n^m\;\hbox{and}\;M^{ijm}\!=\!\epsilon^{ijm}\,$, into the generic rank-2 Killing tensor (\ref{Exp3}).
Inserting (\ref{RL1}) into the second-order constraint of (\ref{RLV}), we get
\beq
\vC(\vx,\vpsi)= q\frac{\vn\times\vx}{r}+\vC(\vpsi\,)\,.
 \label{RL2}
\eeq
In order to solve the first-order constraint of (\ref{RLV}) we write the expansion \cite{B-M} in terms of Grassmann variables,
\beq
C(\vx,\,\vpsi)=C(\vx)+\sum_{k\geq 1}C_{a_{1}\cdots a_{k}}(\vx)\psi^{a_{1}}\cdots\psi^{a_{k}}\,.\label{Expansion}
\eeq
Consequently, the first- and the zeroth-order equations of (\ref{RLV}) can be classified order-by-order in Grassmann-odd variables. Thus, inserting  (\ref{RL2}) in the first-order equation, and  requiring again the vanishing of the commutator,
\beq
\left[\p_{i},\,\p_{j}\right]C(\vx)=0\;\Longrightarrow\;\Delta\left( V(r)-\frac{q^2}{2r^2}\right)=0\,,
\label{Laplace}
\eeq
we deduce the most general radial potential admitting a conserved Laplace-Runge-Lenz vector in the fermion-monopole interaction, namely
\beq
V(r)=\frac{q^2}{2r^2}+\frac{\mu}{r}+\gamma\,,\quad\mu\,,\gamma\in\IR\,.\label{Potential}
\eeq
We can now find the first term in the r.h.s of (\ref{Expansion}), $\,\displaystyle{C(\vx)=\mu\frac{(\vn\cdot\vx)}{r}}\,$. Introducing (\ref{RL2}) and (\ref{Potential}) into the first-order constraint of (\ref{RLV}) leads to $\,\displaystyle{\vC(\vpsi\,)=-\frac{\gyro}{2}\vn\times\vS}\,$ and
\beq\begin{array}{cc}\displaystyle{
\sum_{k\geq 1}C_{a_{1}\cdots a_{k}}(\vx)\psi^{a_{1}}\cdots\psi^{a_{k}}=-\frac{e\gyro}{2}\left(\vS\cdot\vB\right)\left(\vn\cdot\vx\right)-\frac{\gyro q}{2}\left(1-\frac{\gyro}{2}\right)\frac{\vn\cdot\vS}{r}+C(\vpsi)}\,,\\[8pt]
\hbox{with}\quad\gyro(\gyro-4)=0\,.
\end{array}
\label{RL3}
\eeq
The zeroth-order consistency condition of (\ref{RLV}) is only satisfied for $\,\displaystyle{C(\vpsi)=\frac{\mu}{q}\vS\cdot\vn}\,$.  Collecting our results, (\ref{RL1}), (\ref{RL2}), (\ref{Potential}) and (\ref{RL3}), we get a conserved Runge-Lenz vector
if and only if
\beq
\gyro=0\qquad\hbox{or}\qquad\gyro=4\,;
\label{g04}
\eeq
we get namely
\beq
\vK_{\gyro}=\vPi\times\vJ+\mu\,\hx+\left(1-\frac{\gyro}{2}\right)\vS\times\vPi-\frac{e\gyro}{2}\left(\vS\cdot\vB\right)\vx-\frac{\gyro q}{2}\left(1-\frac{\gyro}{2}\right)\frac{\vS}{r}+\frac{\mu}{q}\vS\,.
\eeq

Note that the spin angular momentum which generates the extra ``spin'' symmetry for vanishing gyromagnetic ratio is not more separately conserved for $\gyro=4$. Then, an interesting question is to know if the extra ``spin'' symmetry of $\gyro=0$
is still present for the anomalous superpartner $\gyro=4\,$, cf. section \ref{S6}, in some ``hidden'' way.

Let us consider the ``spin'' transformation generated by the rank-2 Killing tensor with the property,
\beq
C^{mk}(\vx,\vpsi)=2\d^{mk}\big(\vS\cdot\vn\big)-\frac{g}{2}\big(S^{m}n^{k}+S^{k}n^{m}\big)\,,\label{KT2}
\eeq
The previous rank-$2$ Killing tensor, $\,C^{mk}=C^{mk}_{+}\,+\,C^{mk}_{-}\,$, cf. (\ref{KT2}), is obtained by putting
\begin{eqnarray*}
N^{jk}_{+}=({g}/{2})\epsilon^{\;jk}_{l}\,n^l,
\quad
&\widetilde{N}^{jk}_{+\,\;a}=-({i}/{2})\epsilon^{jk}_{\;\,m}\,\epsilon^{m}_{\;\,a_1 a_2},
\\[6pt]
 N^{jkl}_{-}=\big(1-({g}/{2})\big)\epsilon^{jkl},
\quad
&\widetilde{N}^{jkl}_{-\;\,a}=-({i}/{4})\epsilon^{jkl}\,n_{m}\,\epsilon^{m}_{\;\,a_1 a_2}\,
\end{eqnarray*}
 into the general rank-2 Killing tensor (\ref{Exp3}). Inserting (\ref{KT2}) into the second-order constraint of (\ref{RLV}) provides us with
\beq
\vC(\vx,\vpsi)=-\frac{qg}{2}\frac{\big(\vS\times\vn\big)}{r}+\vC(\psi)\quad\hbox{and}\quad g(g-4)=0\,.
\eeq
We use the potential  (\ref{Potential}) to solve the first-order equation of (\ref{RLV}),
\beq\begin{array}{ll}\displaystyle{
C(\vx,\vpsi)=\left(2V(r)-\frac{q^{2}g^{2}}{8r^{2}}-\frac{\mu g^2}{4r}\right)\vS\cdot\vn+c(\psi)}\,,
\\[10pt]\displaystyle{
\vC(\psi)=\frac{\mu g}{2q}\vn\times\vS\qquad\hbox{and}\qquad\gyro\big(\gyro-4\big)=0}\,.
\end{array}
\eeq
The zeroth-order consistency condition is satisfied with $\displaystyle{c(\psi)=-\frac{\gyro^{2}}{8}\frac{\mu^{2}}{q^{2}}\vS\cdot\vn\,,}$ so that collecting our results leads to the conserved vector,
\beq\begin{array}{ll}\displaystyle{
\vo_{\gyro}=\left(\vPi^{2}+\big(2-\frac{\gyro^{2}}{4}\big)V(r)\right)\vS-\frac{\gyro}{2}\big(\vPi\cdot\vS\big)\vPi+\frac{\gyro}{2}\big(\frac{q}{r}+\frac{\mu}{q}\big)\vS\times\vPi}\\[8pt]
\displaystyle{
-\frac{\gyro^{2}}{4}\big(\frac{\mu^{2}}{2q^{2}}-\gamma\big)\vS}\quad\displaystyle{\hbox{with}\quad\gyro\big(\gyro-4\big)=0}\,.
\end{array}
\eeq
In conclusion, the additional $\,o(3)_{spin}\,$ ``spin'' symmetry is recovered in the same particular cases of anomalous gyromagnetic ratios $0$ and $4$ cf. (\ref{g04}). 

$\bullet$ For $\gyro=0$, in particular,
\beq
\vo_{0}=2\E\,\vS\,.
\eeq

$\bullet$ For $\gyro=4$, we find an expression equivalent to that of D'Hoker and Vinet \cite{DV2}, namely 
\beq
\vo_{4}=\left(\vPi^{2}-2V(r)\right)\vS-2\big(\vPi\cdot\vS\big)\,\vPi+2\big(\frac{q}{r}+\frac{\mu}{q}\big)\vS\times\vPi-4\left(\frac{\mu^{2}}{2q^{2}}-\gamma\right)\vS\,.
\eeq
Note that this extra symmetry is generated by a
\emph{Killing tensor}, rather than a Killing vector, as for ``ordinary'' angular momentum. Thus, for sufficiently low energy, the motions are bounded and the conserved vectors $\vJ,\,\vK_{\gyro}$ and $\vo_{\gyro}\,$ generate an $o(4)\oplus o(3)_{spin}\,$ bosonic symmetry algebra.

\section{$\N=2$ Supersymmetry of the fermion-monopole system}\label{S6}
\noindent
So far we have seen that, for a spinning particle with a single Grassmann variable, SUSY and dynamical symmetry are inconsistent, since they require different values for the $\gyro$-factor. Now, adapting the idea of D'Hoker and Vinet to our framework, we show that the
two contradictory conditions can be conciliated by doubling the odd degrees of freedom. The
systems with $\gyro=0$ and $\gyro=4$ will then become
superpartners inside a unified system \cite{FH}.

We consider, hence, a charged  spin-$\frac{1}{2}$ particle moving in a flat manifold $\,\M^{D+2d}\,$, interacting with a static magnetic field $\,\vB\,$. The fermionic degrees of freedom are now carried by a $2d$-dimensional internal space. This is to be compared with the $d$-dimensional internal space sufficient to describe the $\,\N=1\,$ SUSY of the monopole. In terms of Grassmann-odd variables $\,\psi_{1,2}\;$, the local coordinates of the fermionic extension $\,\M^{2d}\,$ read $\left(\psi^{a}_{1},\,\psi^{b}_{2}\right)$ with $\,a,b=1,\cdots,d\,$. The system is still described by the Pauli-like Hamiltonian (\ref{Hamiltonian}). Choosing $\,d=3\,$, we consider the fermion $\xi_{\a}\,$ which is a two-component spinor, $\,\xi_{\a}=\left(\begin{array}{c} \psi_{1}\\\psi_{2}\end{array}\right)\,$, and whose conjugate is $\bar{\xi}^{\a}\,$. Thus, we have a representation of the spin angular momentum,
\beq
S^{k}=\frac{1}{2}\bar{\xi}^{\a} \,\sigma^{k\;\,\b}_{\,\a}\,\xi_{\b}\quad\hbox{with}\quad\a,\b=1,2\,,
\eeq 
and the $\,\sigma^{k\;\,\b}_{\,\a}\,$ with $\,k=1,2,3\,$ are the standard Pauli matrices. Defining the covariant Poisson-Dirac brackets as
\beqa
\big\{f,h\big\}&=&\p_j f\,\frac{\p h}{\p \Pi_j}-\frac{\p f}{\p \Pi_j}\,\p_j h 
+e\,\epsilon_{ijk}B^k\,\frac{\p f}{\p \Pi_i}\frac{\p h}{\p \Pi_j}
+i(-1)^{a^{f}}\left(\frac{\p f}{\p \xi_{\a}}\frac{\p h}{\p\bar{\xi}^{\a}}+\frac{\p f}{\p \bar{\xi}^{\a}}\frac{\p h}{\p\xi_{\a}}\right),\quad
\eeqa
we deduce the non-vanishing fundamental brackets,
\beqa\begin{array}{ll}
\big\{x^{i},\Pi_{j}\big\}=\d^{i}_{j},\quad\big\{\Pi_{i},\Pi_{j}\big\}=e\,\epsilon_{ijk}B^k,\quad\big\{\xi_{\a},\bar{\xi}^{\b}\big\}=-i\d^{\;\b}_{\a},\\[8pt]\displaystyle{
\big\{S^{k},S^{l}\big\}=\epsilon^{kl}_{\;\;m}S^{m},\quad\big\{S^{k},\bar{\xi}^{\b}\big\}=-\frac{i}{2}\bar{\xi}^{\mu}\sigma^{k\;\,\b}_{\,\mu},\quad\big\{S^{k},\xi_{\b}\big\}=\frac{i}{2}\sigma^{k\;\,\nu}_{\,\b}\xi_{\nu}}\,.
\end{array}
\eeqa
We also introduce an auxiliary scalar field, $\,\Phi(r)\,$, satisfying \textit{the ``self-duality'' or
`` Bogomolny'' relation}\footnote{See \cite{FH} to justify terminology.},
\beq
\big\{\Pi^{k},\Phi(r)\big\}=\pm eB^{k}\,.\label{SelfDuality}
\eeq
This auxiliary scalar field also defines a square root of the external potential of the system so that $\,\displaystyle{\frac{1}{2}\Phi^{2}(r)=V(r)}\,$. As an example we obtain the potential \footnote{The constant is $\,\displaystyle{\gamma=\frac{\mu^2}{2q^2}}\,$.} in (\ref{Potential}) by considering the auxiliary field, $\,\displaystyle{\Phi(r)=\pm\left(\frac{q}{r}+\frac{\mu}{q}\right)}\,$. 

In order to investigate the $\,\N=2\,$ supersymmetry of the Pauli-like Hamiltonian (\ref{Hamiltonian}), we outline the algorithm developed we use to construct supercharges linear in the gauge covariant momentum,
\beqa\left\lbrace \begin{array}{lll} 
\displaystyle{\mp e\Phi(r)\,B^{j}C^j+\frac{ie\gyro}{4}B^{k}\left(\bar{\xi}^{\mu}\sigma^{k\,\nu}_{\mu}\frac{\p C}{\p\bar{\xi}^{\nu}}-\frac{\p C}{\p\xi_{\mu}}\sigma^{k\,\nu}_{\mu}\xi_{\nu}\right)
-\frac{e\gyro}{4}\,\bar{\xi}^{\mu}\sigma^{k\,\nu}_{\mu}\xi_{\nu}\,C^{j}\p_{j}B^{k}
=0}\,,&\hbox{order 0}& 
\\[10pt]
\displaystyle{
\p_{m}C=e\,\epsilon_{mjk}B^kC^j+i\frac{e\gyro}{4}B^k\left(\bar{\xi}^{\mu}\sigma^{k\,\nu}_{\mu}\frac{\p C^m}{\p\bar{\xi}^{\nu}}-\frac{\p C^m}{\p\xi_{\mu}}\sigma^{k\,\nu}_{\mu}\xi_{\nu}\right) } \,, &\hbox{order 1}& 
\\[12pt]
\p_jC^k(x,\xi,\bar{\xi})+\p_kC^j(x,\xi,\bar{\xi})=0&\hbox{order 2}\,.&
\label{2susy}
\end{array}\right.
\eeqa
Let us first consider the Killing spinor,
\beq
C_{\b}^j=\frac{1}{2}\sigma^{j\,\a}_{\b}\,\xi_{\a}\,.
\eeq
Inserting this Killing spinor into the first-order equation of (\ref{2susy}) provides us with
\beq
\p_m C_{\b}=-\frac{i}{2}e\,B_m\,\xi_{\b}\quad\hbox{and}\quad\gyro=4\,,
\eeq
which can be solve using \textit{the  self-duality relation (\ref{SelfDuality})}. We get $\,\displaystyle{C_{\b}(\vx,\vxi)=\pm\frac{i}{2}\Phi(r)\,\xi_{\b}}\,$, provided the anomalous gyromagnetic factor is $\,g=4\,$. The zeroth-order constraint of (\ref{2susy}) is identically satisfied, so that collecting our results provides us with the supercharge,
\beq
\Q_{\b}=\frac{1}{2}\Pi_j\,\sigma^{j\,\a}_{\b}\,\xi_{\a}\pm\frac{i}{2}\Phi(r)\xi_{\b}\,.
\label{2susy2}
\eeq
To obtain the supercharge conjugate to (\ref{2susy2}), we consider the Killing spinor,
\beq
\bar{C}^{k\,\b}=\frac{1}{2}\bar{\xi}^{\a}\,\sigma_{\a}^{k\,\b}\,.
\eeq
We solve the first-order equation of (\ref{2susy}) for the anomalous value of the gyromagnetic ratio $\,g=4\,$ using the Bogomolny equation (\ref{SelfDuality}). This leads to the conjugate $\displaystyle{\bar{C}^{\b}(\vx,\vxi)=\mp\frac{i}{2}\Phi(r)\bar{\xi}^{\b}}$. The zeroth-order consistency constraint is still satisfied and we obtain the odd-supercharge,
\beq
\bar{\Q}^{\b}=\frac{1}{2}\bar{\xi}^{\a}\,\sigma_{\a}^{k\,\b}\Pi_{k}\mp\frac{i}{2}\Phi(r)\,\bar{\xi}^{\b}\,.
\label{2susy3}
\eeq
The supercharges $\,\Q_{\b}\,$ and $\,\bar{\Q}^{\b}\,$ are, both, square roots of the Pauli-like Hamiltonian $\,\H_{4}\,$ and therefore \textit{generate the $\N=2$ supersymmetry of the spin-monopole field system,}
\beq
\big\{\bar{\Q}^{\b},\Q_{\b}\big\}=-i\H_{4}\,\Id\,.
\eeq
It is worth noting that defining the rescaled, $\,\displaystyle{\bar{\U}^{\b}=\bar{\Q}^{\b}\frac{1}{\sqrt{\H_4}}}\,$ and $\,\displaystyle{\U_{\b}=\frac{1}{\sqrt{\H_4}}\Q_{\b}}\,$, it is straightforward to get,
\beq\displaystyle{
\H_0=\bar{\U}^{\b}\,\H_4\,\U_{\b}}\,,
\eeq
which make manifest the fact that the two anomalous cases $\,\gyro=0\,$ and $\,\gyro=4\,$ can be viewed as superpartners \footnote{With The scalar $\,\bar{\xi}^{\b}\xi_{\b}=2\,$.}, cf \cite{FH}. Moreover, in our enlarged system, the following bosonic charges
\beqa\begin{array}{lll}\displaystyle{
\vJ=\vx\times\vPi-q\,\hx+\vS}\,,\\[10pt]\displaystyle{
\vK=\vPi\times\vJ+\mu\,\hx-\vS\times\vPi-2e\left(\vS\cdot\vB\right)\vx+2q\frac{\vS}{r}+\frac{\mu}{q}\vS}\,,\\[14pt]\displaystyle{
\vo=\bar{\Q}^{\b}\,\vsigma_{\b}^{\;\a}\,\Q_{\a}=\frac{1}{2}\left(\Phi^2(r)-\vPi^{2}\right)\vS+\big(\vPi\cdot\vS\big)\vPi\mp\Phi(r)\,\vS\times\vPi,}
\end{array}
\eeqa
remain conserved such that they form, together with the supercharges $\,\Q_{\b}\,$ and $\,\bar{\Q}^{\b}\,$, the classical symmetry superalgebra \cite{DV2,FH},
\beqa\begin{array}{llll}\displaystyle{
\big\{\bar{\Q}^{\b},\Q_{\b}\big\}= -i\H_4\,\Id\,,\quad\big\{\bar{\Q}^{\b},\bar{\Q}^{\b}\big\}=\big\{\Q_{\b},\Q_{\b}\big\}=0\,,\quad\big\{\bar{\Q}^{\b},J^k\big\}=\frac{i}{4}\bar{\Q}^{\a}\sigma^{k\,\b}_{\a}}\,,\\[12pt]\displaystyle{
\big\{\Q_{\b},J^k\big\}=-\frac{i}{4}\sigma^{k\,\a}_{\b}\Q_{\a}\,,\quad\big\{\bar{\Q}^{\b},K^j\big\}=-\frac{i}{4}\,\frac{\mu}{q}\,\bar{\Q}^{\a}\sigma^{j\,\b}_{\a}\,,\quad\big\{\Q_{\b},K^j\big\}=\frac{i}{4}\,\frac{\mu}{q}\,\sigma^{j\,\a}_{\b}\Q_{\a}}\,,\\[12pt]\displaystyle{
\big\{\bar{\Q}^{\b},\Omega^k\big\}=-i\H_4\,\bar{\Q}^{\a}\sigma^{k\,\b}_{\a}\,,\quad\big\{\Q_{\b},\Omega^k\big\}=i\H_4\,\sigma^{k\,\a}_{\b}\Q_{\a}\,,\quad\big\{\Omega^i,K^j\big\}=\frac{\mu}{q}\epsilon^{ijk}\,\Omega^k}\,,\\[12pt]\displaystyle{
\big\{K^i,K^j\big\}=\epsilon^{ijk}\left[\left(\frac{\mu^2}{q^2}-2\H_4\right)J^k+2\Omega^k\right]\,,\quad\big\{\Omega^i,\Omega^j\big\}=\epsilon^{ijk}\,\H_4\,\Omega^k}\,,\\[12pt]\displaystyle{
\big\{J^i,\Lambda^j\big\}=\epsilon^{ijk}\Lambda^k\quad\hbox{with}\quad\Lambda^l=J^l,K^l,\Omega^l}\,.
\end{array}\nonumber
\eeqa

\section{{\bf Planar System}}\label{S7}

\noindent
In 2 dimensions the models simplify. The magnetic field is $F_{ij} = \varepsilon_{ij} B = \p_i A_j - \p_j A_i$ and
the spin tensor is actually a scalar
\beq
S = - \frac{i}{2}\, \varepsilon_{ij} \psi_i \psi_j.
\label{a.2}
\eeq
The Hamiltonian takes the form
\beq
H = \frac{1}{2}\, \vec{\Pi}^2 - \frac{e\gyro}{2}\, S B + V(r).
\label{a.1}
\eeq
The fundamental brackets remain the same as in (\ref{PBrackets}).
The dynamical quantities (\ref{Exp}) become constants of motion if the constraints (\ref{constraints}) are satisfied:
\beqa\begin{array}{lll}
\displaystyle{ C_i \p_i H + i \frac{\p H}{\p \psi_i} \frac{\p C}{\p \psi_i} = 0}\,,&\hbox{order 0}&\\[14pt]

\displaystyle{ \p_i C = e F_{ij} C_j + i \frac{\p H}{\p \psi_j} \frac{\p C_i}{\p \psi_j} + C_{ij} \p_j H }\,,&\hbox{order 1}&\\[14pt]

\displaystyle{ \p_i C_j + \p_j C_i = e \left( F_{ik} C_{kj} - C_{ik} F_{kj} \right) + i \frac{\p H}{\p \psi_k} \frac{\p C_{ij}}{\p \psi_k}
 + C_{ijk} \p_k H}\,,&\hbox{order 2}&\\[14pt]
 
 \p_i C_{jk} + \p_j C_{ki} + \p_k C_{ij} = C_{ijkl} \p_l H + (\mbox{terms linear in $C_{lmn}$})&\hbox{order 3}&.\end{array}
\eeqa
Using
\beq
i \frac{\p H}{\p \psi_i} = - \frac{e\gyro}{2}\, F_{ij} \psi_j = - \frac{e\gyro}{2}\, B \varepsilon_{ij} \psi_j,
\label{a.5}
\eeq
the first (zeroth-order) constraint becomes
\beq
\frac{e\gyro}{2}\, B \varepsilon_{ij} \psi_j \frac{\p C}{\p \psi_i} = C_i \left(\p_i V - \frac{e\gyro}{2}\, S\, \p_i B \right),
\label{a.4}
\eeq
complemented by the first-order equation
\beq
\p_i C = eB \left( \varepsilon_{ij} C_j + \frac{\gyro}{2}\, \varepsilon_{jk} \psi_j \frac{\p C_i}{\p \psi_k} \right)
 + C_{ij} \left(  \p_j V - \frac{e\gyro}{2}\, S  \p_j B \right).
\label{a.7}
\eeq
Similarly the second and higher-order equations take the form
\beq
\p_i C_j + \p_j C_i = eB \left( \varepsilon_{ik} C_{kj} + \varepsilon_{jk} C_{ki} + \frac{\gyro}{2}\,\varepsilon_{jk} \psi_j \frac{\p C_i}{\p \psi_k} 
 \right) + C_{ijk} \left(  \p_k V - \frac{e\gyro}{2}\, S  \p_k B \right),
\label{a.8}
\eeq
etc. For radial functions $V(r)$ and $B(r)$:
$
\p_i V = (x_i/r)\,V^{\prime},\
 \p_i B = (x_i/r)\,B^{\prime}\,,
$
hence
\beq
C_{i...j} \left(  \p_j V - \frac{e\gyro}{2}\, S  \p_j B \right) = \frac{C_{i...j}\, x_j}{r} \left( V^{\prime} - \frac{e\gyro}{2}\, S B^{\prime} \right).
\label{a.9}
\eeq
Let us now consider some specific cases. Universal generalized Killing vectors are
\beq
C_i = (\gamma_i,\, \varepsilon_{ij} x_j,\,  \psi_i,\, \varepsilon_{ij} \psi_j),
\label{a.10}
\eeq
with $\gamma_i$ a constant vector. Observe that $S$ is a constant of motion itself:
\beq
\left\{ H, S \right\} = 0,
\label{a.11}
\eeq
and all quantities quadratic in the Grassmann variables are proportional to $S$. 

\vskip2mm
$\bullet$
A constant Killing vector $\gamma_i$ gives a constant of motion only if we can find solutions for the equations
\beq
\p_i C = e B \varepsilon_{ij} \gamma_j, \hspace{2em} 
B  \varepsilon_{ji} \psi_i \frac{\p C}{\p \psi_j} = \gamma_i \left( \frac{2}{e\gyro}\, \p_i V- S \p_i B \right).
\label{a.a1}
\eeq
Now for a Grassmann-even function $C = c_0 + c_2 S$ the left-hand side of the second equation vanishes, therefore we
must require $B$ and $V$ to be constant. Taking $V = 0$, this leads to the solution
\beq
C = - e B \varepsilon_{ij} \gamma_i x_j, \hspace{2em} V = 0, \hspace{2em} B = \mbox{constant}.
\label{a.a2}
\eeq
The corresponding constant of motion is $\gamma_i P_i$, with
\beq
P_i = \Pi_i - e B \varepsilon_{ij} x_j,
\label{a.a3}
\eeq
identified with \textit{``magnetic translations''} \cite{Kostel}.
\vskip2mm
$\bullet$
Next we consider the linear Killing vector $C_i = \varepsilon_{ij} x_j$, with all higher-order coefficients $C_{ij...} = 0$. 
Again for Grassmann-even $C$ the left-hand side of eq.\ (\ref{a.4}) vanishes, and we get the condition
\beq
\varepsilon_{ij} x_i \p_j B = \varepsilon_{ij} x_i \p_j V = 0,
\label{a.a4}
\eeq
which is automatically satisfied for radial functions $B(r)$ and $V(r)$. Therefore we only have to solve eq.\ (\ref{a.7}): 
\beq
\p_i C = - e B  x_i = - \frac{e x_i}{r}\, (rB). 
\label{a.a5}
\eeq
We infer that $C(r)$ is a radial function, with
$
C^{\prime} = -e rB.
$
Therefore $C$ is given by \textit{the magnetic flux through the disk $D_r$ centered at the origin with radius $r$}:
\beq
C = - \frac{e}{2\pi}\, \int_{D_r} B(r) d^2x \equiv - \frac{e}{2\pi}\, \Phi_B(r).
\label{a.a7}
\eeq
We then find the constant of motion representing angular momentum:
\beq
L = \varepsilon_{ij} x_i \Pi_j +\frac{e}{2\pi}\,  \Phi_B(r).
\label{a.a8}
\eeq

$\bullet$
There are two Grassmann-odd Killing vectors, the first one being $C_i = \psi_i$. With this Ansatz, we get for the 
scalar contribution to the constant of motion the constraints
\beq
\frac{e\gyro}{2} B\, \varepsilon_{ij} \psi_j\, \frac{\p C}{\p \psi_i} = \psi_i \p_i V\,, \hspace{2em}
\p_i C = \frac{eB}{2} \left( 2 - \gyro \right) \varepsilon_{ij} \psi_j\,.
\label{a.b1}
\eeq
It follows that either $\gyro = 2$ and $(C, V)$ are constant (in which case one may take $C = V = 0$),  
or $ \gyro \neq 2$ and $C$ is of the form
\beq
C = \varepsilon_{ij} K_i(r) \psi_j\,
\quad\hbox{with}\quad 
\p_i V = - \frac{e\gyro}{2}\, B K_i\,, \hspace{2em} 
\p_i K_j = \frac{(2-\gyro)eB}{2}\, \delta_{ij}.
\label{a.b3}
\eeq
This is possible only if $B$ is constant and
\beq
K_i = \frac{eB(2-\gyro)}{2}\,x_i \equiv \kappa x_i\,, \hspace{2em} 
V(r) = \frac{\gyro(\gyro-2)}{8}\, e^2 B^2 r^2 = - \frac{e\gyro \kappa}{4\pi}\, \Phi_B(r).
\label{a.b4}
\eeq
It follows that we have a conserved supercharge of the form \cite{Kostel}
\beq
Q = \psi_i \left( \Pi_i - \kappa \varepsilon_{ij} x_j \right).
\label{a.b5}
\eeq
The bracket algebra of this supercharge takes the form
\beq
i \left\{ Q, Q \right\} = 2 H + (2 - g)e B J, \hspace{2em} J = L + S.
\label{a.b6}
\eeq
Of course, as $S$ and $L$ are separately conserved, $J$ is a constant of motion as well.
\vspace{1ex}

$\bullet$
Finally we consider the dual Grassmann-odd Killing vector $C_i = \varepsilon_{ij} \psi_j$. Then the 
constraints (\ref{a.4}) and (\ref{a.7}) become
\beq
\frac{eg}{2}\,B\,\frac{\p C}{\p \psi_i} = \p_i V, \hspace{2em}
\p_i C = \frac{(g-2)eB}{2}\, \psi_i, 
\label{a.c1}
\eeq
implying that $C = N_i(x) \psi_i$ and
\beq
\frac{eg}{2}\, B\, N_i = \p_i V, \hspace{2em}
\p_i N_j = \frac{(g-2) eB}{2}\, \delta_{ij}.
\label{a.c2}
\eeq
As before, $B$ must be constant and the potential is identical to (\ref{a.b4}):
\beq
N_i = - \kappa x_i, \hspace{2em} 
V = - \frac{eg\kappa}{4\pi}\, \Phi_B(r) = \frac{g(g-2)}{8}\, e^2 B^2 r^2.
\label{a.c3}
\eeq
Thus we find the dual conserved supercharge \cite{HmonRev},
\beq
\tilde{Q} = \varepsilon_{ij} \psi_i \left( \Pi_j - \kappa \varepsilon_{jk} x_k \right) 
 = \psi_i \left( \varepsilon_{ij} \Pi_j + \kappa x_i \right),
\label{a.c4}
\eeq
which satisfies the bracket relations
\beq
i \left\{ \tilde{Q}, \tilde{Q} \right\} = 2 H + (2-g) eB J, \hspace{2em}
i \left\{ Q, \tilde{Q} \right\} = 0.
\label{a.c5}
\eeq
Thus the harmonic potential (\ref{a.b4}) with constant magnetic field $B$ allows a classical $\N = 2$ supersymmetry 
with supercharges $(Q, \tilde{Q})$, whilst the special conditions $g = 2$ and $V = 0$ allows for $\N = 2$
supersymmetry for any $B(r)$.

\section{Discussion}\label{S8}
\noindent
In this paper we studied, in the framework of the covariant Hamiltonian dynamics proposed in \cite{vH}, the symmetries and the supersymmetries of a spinning particle coupled to a magnetic monopole field. The gyromagnetic ratio  determines the type of (super)symmetry the system can admit~: for the Pauli-like hamiltonian (\ref{Hamiltonian}) $\,\N=1\,$ supersymmetry   only arises for gyromagnetic ratio $\gyro=2$ and with no potential, $V=0$, confirming  Spector's observation \cite{Spector}. We also derived additional supercharges, which are not square roots of the Hamiltonian of the system, though.

A Runge-Lenz-type dynamical symmetry  requires instead an anomalous gyromagnetic ratio,
$$\gyro=0\quad\hbox{or}\quad\gyro=4\,,$$
with the additional bonus of an extra ``spin" symmetry.
These particular values of $\gyro$ come from the effective coupling of the form $F_{ij}\mp\epsilon_{ijk}D_k\Phi$, which
add or cancel   for self-dual fields, $F_{ij}=\epsilon_{ijk}D_k\Phi$  \cite{FH}.

The super- and the bosonic symmetry can be combined; the price to pay is, however, to enlarge the fermionic space, as proposed by D'Hoker and Vinet \cite{DV2} (see also \cite{FH}); this provides us with an $\,\N=2\,$ SUSY.

Our recipe  also applies to a planar  fermion in any  planar magnetic field [i.e. one perpendicular to the plane]. As an illustration, we have shown, for ordinary gyromagnetic, that in addition to the usual supercharge (\ref{a.b5}) generating the supersymmetry,  the system also admits another square root of the Pauli Hamiltonian $\,H\,$ \cite{HmonRev}. This happens due to the existence
of a dual Killing tensor.

At last, we remark that confining the spinning particle onto a sphere of fixed radius $\,\rho\,$ implies the set of constraints \cite{DJ-al},
\beq
\vx^2=\rho^2\,,\quad\vx\cdot\vpsi=0\quad\hbox{and}\quad\vx\cdot\vPi=0\,.
\eeq  
This freezes the radial potential to a constant,
and we recover the $\,\N=1\,$ SUSY  described by the supercharges $\,\Q\,$, $\,\Q_1\,$ and $\,\Q_2\,$ for ordinary gyromagnetic factor $\,\gyro=2\,$.


\begin{acknowledgments}\noindent
One of us (JPN) is indebted, to the {\it R\'egion Centre} for a doctoral scholarship, to the {\it Laboratoire de Math\'ematiques et de Physique Th\'eorique} of Tours University and to the {\it NIKHEF (Amsterdam)}  for hospitality extended to him. 
\end{acknowledgments}


\end{document}